\documentclass[preprint, 11pt]{elsarticle}

\usepackage{fullpage}
\usepackage{multirow}
\usepackage{colordvi}
\usepackage{color}
\usepackage{booktabs}
\usepackage{epsfig}
\usepackage{amsmath}
\usepackage{amssymb}
\usepackage{url}

\usepackage{xspace}
\newcommand{\RAMm}{RAM\xspace}
\newcommand{\FSAX}{\texttt{f\_a\_times\_x\_}\xspace}
\newcommand{\FSATY}{\texttt{f\_at\_times\_y\_}\xspace}
\newcommand{\FCAAT}{\texttt{f\_get\_col\_aat\_}\xspace}
\newcommand{\FDAAT}{\texttt{f\_get\_diag\_aat\_}\xspace}
\newcommand{\gprof}{\texttt{gprof}\xspace}

\newcommand{\specialcell}[2][c]{\begin{tabular}[#1]{@{}c@{}}#2\end{tabular}}

\newcommand{\ignore}[1]{}

\newtheorem{lemma}{Rule}

\def\Problem#1{{\tt #1}}

\journal{Journal of Computational and Applied Mathematics}

\begin{document}

\begin{frontmatter}

\title{Solving Large-Scale Optimization Problems Related to Bell's Theorem}

\author[edi]{Jacek Gondzio}
\ead{J.Gondzio@ed.ac.uk}
\ead[url]{http://www.maths.ed.ac.uk/~gondzio/}

\author[edi,iftia]{Jacek A. Gruca\corref{cor1}}
\ead{jacek.a@gruca.org}
\ead[url]{http://www.gruca.org/}
\cortext[cor1]{Corresponding author. Tel. +48 602153844; fax +48 585232056. 
Postal address: Institute of Theoretical Physics and Astrophysics, 
University of Gda\'{n}sk, 80-952 Gda\'{n}sk, ul. Wita Stwosza 57, Poland}

\author[edi]{J. A. Julian Hall}
\ead{J.A.J.Hall@ed.ac.uk}
\ead[url]{http://www.maths.ed.ac.uk/hall/}

\author[iftia]{Wies{\l}aw Laskowski}
\ead{wieslaw.laskowski@univ.gda.pl}

\author[iftia]{Marek \.{Z}ukowski}
\ead{marek.zukowski@univie.ac.at}

\address[edi]{School of Mathematics and Maxwell Institute for Mathematical Sciences, The University of Edinburgh,\\Mayfield Road, Edinburgh EH9 3JZ, United Kingdom}

\address[iftia]{Institute of Theoretical Physics and Astrophysics, University of Gda\'{n}sk,\\80-952 Gda\'{n}sk, Poland}

\begin{abstract}
Impossibility of finding local realistic models for quantum correlations due to entanglement is an important fact in foundations of quantum physics, gaining now new applications in quantum information theory. We present an in-depth description of a method of testing the existence of such models, which involves two levels of optimization: a higher-level non-linear task and a lower-level linear programming (LP) task. The article compares the performances of the existing implementation of the method, where the LPs are solved with the simplex method, and our new implementation, where the LPs are solved with an innovative matrix-free interior point method. We describe in detail how the latter can be applied to our problem, discuss the basic scenario and possible improvements and how they impact on overall performance. Significant performance advantage of the matrix-free interior point method over the simplex method is confirmed by extensive computational results. The new method is able to solve substantially larger problems. Consequently, the noise resistance of the non-classicality of correlations of several types of quantum states, which has never been computed before, can now be efficiently determined. An extensive set of data in the form of tables and graphics is presented and discussed. The article is intended for all audiences, no quantum-mechanical background is necessary.
\end{abstract}


\end{frontmatter}

%%%%%%%%%%%%%%%%%%%%%%%%%%%%%%%%%%%%%%%%%%%%%
\section{Introduction}
\label{intro}

We discuss a new approach to tackling a certain class of large scale optimization problems arising in quantum information science and linked with Bell's Theorem \cite{mybib:Bell}. These problems were described in detail in \cite{mybib:Kaszlikowski, mybib:GrucaEtAl} and are, in fact, two-level optimization problems. The higher level problem is a non-convex non-linear optimization task. It requires the solution of a sequence of linear programming problems (LPs for short). Each of the LPs is a lower level task. We will focus on LP, as it has been identified as the major computational challenge. Reference \cite{mybib:GrucaEtAl} reports a range of problems which have been solved, with the small and medium scale LPs being solved using the GLPK~\cite{mybib:GLPK} simplex implementation. However, even for medium scale problems the CPU times were occasionally prohibitive. The solution time of a single problem with about 64,000 variables using GLPK sometimes exceeds 24 hours and the upper level optimization may require the solution of hundreds of such LP problems. 

The solution of a single LP is a clear bottleneck in quantum information optimization. 
In this paper we report on the efforts to accelerate this part of optimization 
process. We make two contributions to achieve the goal: 
\begin{itemize}
\item
we reformulate the problem by removing multiple redundant constraints from its 
original definition \cite{mybib:GrucaEtAl}; 
\item
we replace the simplex method with a specialized variant of the matrix-free 
interior point method \cite{mybib:G-MtxFree}. 
\end{itemize}
Although eliminating redundancy significantly reduces the number of LP 
constraints and produces more compact problem formulations, it is not 
sufficient to extend the applicability of the previous approach \cite{mybib:GrucaEtAl} 
which was based on the use of the simplex method. The reduced problems 
still defy both {\it standard} \/ LP approaches: the simplex method and 
the interior point method. The use of the matrix-free variant of an interior 
point method provides a major advance because it allows the solution 
of these problems to be faster by orders of magnitude and guarantees 
a major reduction of the memory required to store them.  

The aim of this article is twofold. Firstly, it is to describe the 
advantages of the interior point method over the simplex method in the context of the described problem
and secondly, to include a complete formulation 
of the problem of computing the minimal critical visibility, accessible also 
to researchers outside the quantum information domain. In the following two sections we cover the second aim by discussing the formulation of the problems without going into excessive details on quantum physics in order to make the paper accessible to non-specialists. We then discuss the original approach and the method along with its advantages and drawbacks, then we present our revised approach and a comparison between the two, thus covering the first aim of the article. Finally, we draw some conclusions and suggestions for potential future investigation.

\section{Motivation}
\label{motiv}

{\em Quantum information science}, also called quantum informatics, is an attempt to harness paradoxical aspects of quantum physics in the form of highly non-classical effects in information transfer and processing, aiming at breaking classical limits (in the form of Shannon's information theory or principles of Church-Turing computation). It operates on quantum entangled states of particles and aims to provide protocols which are faster, more robust, secure or in any other way better than the ones provided by classical informatics. In order to do this it must harness the non-classical properties of the states it operates on. This has led to a need for a measure of non-classicality of quantum entangled states. Such a measure is also important in practical applications where obtaining a pure state is impossible since it always comes with a bit of noise which reduces its non-classicality. For our purposes, classicality will be synonymous with local realism or local hidden variable models \cite{mybib:Fine-HV} (as it was shown that any local-realistic approach can be effectively simulated by local hidden variable models). In short, local realism requires that no interaction can occur between particles that exceeds the speed of light, and that particles must have a pre-existing value for any measurement before the measurement is made. We will now present a formulation of the problem of computing a certain type of threshold for non-classicality of quantum correlations.

\section{Problem formulation}
\subsection{Critical visibility}

Let us consider an experiment on a quantum entangled state of two particles emitted from a source in distinct directions. These particles travel towards Alice on one side and Bob on the other. The names Alice and Bob refer to the observers who both perform measurements of observables on their particles. The outcome of any single measurement is either 0 or 1. In the simplest case, Alice and Bob choose from two observables each: $A_1, A_2$ for Alice and $B_1, B_2$ for Bob\footnote{Cases where there are more possible measurement outcomes, more observers and more observables available to the observers can also be solved by our approach but are omitted here for simplicity.}. Additionally, we assume that the observers are miles apart to ensure that their measurements cannot interact with one another without exceeding the speed of light (so that locality is assured).

Quantum mechanics provides us with tools to calculate probabilities of measurement outcomes, which are denoted
\begin{align}
P(r_a, r_b | A_i, B_k), ~ r_a, r_b \in \{0, 1\}; i, k \in \{1, 2\}.
\end{align}
The above is a probability of a situation $(r_a, r_b | A_i, B_k)$ by which we denote that Alice obtains outcome $r_a$ while measuring observable $A_i$ and Bob obtains outcome $r_b$ while measuring observable $B_k$. Under local realism there must also exist an underlying global probability distribution 
\begin{align}
\label{plr}
p_{lr}(a_1, a_2, b_1, b_2), 
\end{align}
where $a_1$ and $a_2$ ($b_1$ and $b_2$) are hypothetical values of measurements performed by Alice (Bob), if she (he) chooses to measure $A_1$ or $A_2$ ($B_1$ or $B_2$). Both Alice and Bob have to choose exactly one observable for their measurements on a given pair of particles. The measurements of, say, $A_1$ and $A_2$ may be quantum-mechanically incompatible so are impossible to measure together (due to partial or full complementarity). Thus quantum formalism does not give us any method of deriving (\ref{plr}). As a matter of fact, quantum mechanics rules out the possibility of the existence of $p_{lr}(a_1, a_2, b_1, b_2)$ and hence is a non-local-realistic theory: it is a non-classical theory. The practical question for quantum information applications is: where is the border between the classical and the non-classical?

If we now assume local realism is correct then the following equations hold%\\
\begin{align}
\label{marginals}
P(r_a, r_b| A_1, B_1) &=  \sum_{a_2, b_2 = 0}^1 p_{lr}(r_a, a_2, r_b, b_2)\nonumber\\
P(r_a, r_b| A_1, B_2) &=  \sum_{a_2, b_1 = 0}^1 p_{lr}(r_a, a_2, b_1, r_b)\nonumber\\
P(r_a, r_b| A_2, B_1) &=  \sum_{a_1, b_2 = 0}^1 p_{lr}(a_1, r_a, r_b, b_2)\nonumber\\
P(r_a, r_b| A_2, B_2) &=  \sum_{a_1, b_1 = 0}^1 p_{lr}(a_1, r_a, b_1, r_b).
\end{align}

It can be shown that for some quantum entangled states the marginal sums (\ref{marginals}) cannot be satisfied. This is contained in Bell's Theorem. We will now discuss a method\footnote{One could define other measures of non-classicality of the correlations, e.g. see \cite{mybib:GrucaB}. However, as our aims are to present the basics behind the computational methods, we shall stick here to the simplest one.} of establishing the degree to which these are violated by some quantum entangled states. Establishing thresholds of non-classicality allows us to show the extent to which certain quantum states are robust to noises or distortion.

If a strongly non-classical state $QM$ violates the above marginal sums (\ref{marginals}) with a quantum-mechanical probability distribution:
\begin{align}
P_{QM}(r_a, r_b | A_i, B_k)
\end{align}
then we can admix to it a fraction $f \in [0, 1]$ of a fully classical state called white noise \cite{mybib:Laskowski} to make it satisfy them. By white noise we understand completely random results (with a classical model of two coin tosses, one by Alice and one by Bob). The resulting state, $QM$ with the noise admixture, will have the following probability distribution
\begin{align}
\label{probability}
P(r_a, r_b | A_i, B_k) = v \big(P_{QM}(r_a, r_b | A_i, B_k) - \frac{1}{4}\big) + \frac{1}{4},
\end{align}
where $v = 1 - f$ is the visibility of the original state $QM$. If we admix $f=1$ of white noise to even the most non-classical state, then we will always satisfy~(\ref{marginals}). On the other hand, if $v=1$ then, for some non-classical states, we will see (\ref{marginals}) violated. For such states there exists a threshold visibility, known as the critical visibility, above which we will see violation of the marginal sums. Let us now state the above as a set of linear equations (constraints) which have to be satisfied.  From (\ref{marginals}) and (\ref{probability}) we have:
\begin{align}
&&\sum_{a_2, b_2 = 0}^1 p_{lr}(r_a, a_2, r_b, b_2) + v(\frac{1}{4} - P(r_a, r_b | A_1, B_1))  = \frac{1}{4},\nonumber\\
&&\sum_{a_2, b_1 = 0}^1 p_{lr}(r_a, a_2, b_1, r_b) +v(\frac{1}{4} - P(r_a, r_b | A_1, B_2) ) = \frac{1}{4},\nonumber\\
&&\sum_{a_1, b_2 = 0}^1 p_{lr}(a_1, r_a, r_b, b_2) + v(\frac{1}{4} - P(r_a, r_b | A_2, B_1))  = \frac{1}{4}, \nonumber\\
&&\sum_{a_1, b_1 = 0}^1 p_{lr}(a_1, r_a, b_1, r_b) + v(\frac{1}{4} - P(r_a, r_b | A_2, B_2))  = \frac{1}{4}, \nonumber\\
&&\sum_{a_1,a_2,b_1,b_2=0}^{1} p_{lr}(a_1,a_2,b_1,b_2) = 1; ~~0\leqslant v \leqslant1; ~~ 0\leqslant p_{lr}(\cdot) \leqslant1, \nonumber\\
\end{align}
where the top four lines form 16 equalities (each pair $r_a, r_b$ accounts for four of them as $r_a, r_b \in \{0, 1\}$).  The orthant defined by the above 16 marginal sums, the probability summation constraint and the bounds on the visibility and probabilities forms the feasible region of our LP. The following cost function
\begin{align}
&&z(p_{lr}(0,0,0,0), p_{lr}(0,0,0,1), p_{lr}(0,0,1,0),  \ldots, p_{lr}(1,1,1,1), v) = v,
\end{align}
has a maximum on the feasible region equal to the critical visibility. If we adopt the traditional formulation of the LP problem:
\begin{align}
\text{max}\quad c^Tx\quad \text{subject to}\quad Ax=b,\quad 0\leqslant x \leqslant1,
\end{align}
then, for the problem stated above, we have:
\begin{align}
A = \left[\begin{array}{ccccccccccccccccc}
1 & 1 & 0 & 0 & 1 & 1 & 0 & 0 & 0 & 0 & 0 & 0 & 0 & 0 & 0 & 0 & 1/4 - P(0, 0, | A_1, B_1)\\
0 & 0 & 1 & 1 & 0 & 0 & 1 & 1 & 0 & 0 & 0 & 0 & 0 & 0 & 0 & 0 & 1/4 - P(0, 1, | A_1, B_1)\\
0 & 0 & 0 & 0 & 0 & 0 & 0 & 0 & 1 & 1 & 0 & 0 & 1 & 1 & 0 & 0 & 1/4 - P(1, 0, | A_1, B_1)\\
0 & 0 & 0 & 0 & 0 & 0 & 0 & 0 & 0 & 0 & 1 & 1 & 0 & 0 & 1 & 1 & 1/4 - P(1, 1, | A_1, B_1)\\
1 & 0 & 1 & 0 & 1 & 0 & 1 & 0 & 0 & 0 & 0 & 0 & 0 & 0 & 0 & 0 & 1/4 - P(0, 0, | A_1, B_2)\\
0 & 1 & 0 & 1 & 0 & 1 & 0 & 1 & 0 & 0 & 0 & 0 & 0 & 0 & 0 & 0 & 1/4 - P(0, 1, | A_1, B_2)\\
0 & 0 & 0 & 0 & 0 & 0 & 0 & 0 & 1 & 0 & 1 & 0 & 1 & 0 & 1 & 0 & 1/4 - P(1, 0, | A_1, B_2)\\
0 & 0 & 0 & 0 & 0 & 0 & 0 & 0 & 0 & 1 & 0 & 1 & 0 & 1 & 0 & 1 & 1/4 - P(1, 1, | A_1, B_2)\\
1 & 1 & 0 & 0 & 0 & 0 & 0 & 0 & 1 & 1 & 0 & 0 & 0 & 0 & 0 & 0 & 1/4 - P(0, 0, | A_2, B_1)\\
0 & 0 & 1 & 1 & 0 & 0 & 0 & 0 & 0 & 0 & 1 & 1 & 0 & 0 & 0 & 0 & 1/4 - P(0, 1, | A_2, B_1)\\
0 & 0 & 0 & 0 & 1 & 1 & 0 & 0 & 0 & 0 & 0 & 0 & 1 & 1 & 0 & 0 & 1/4 - P(1, 0, | A_2, B_1)\\
0 & 0 & 0 & 0 & 0 & 0 & 1 & 1 & 0 & 0 & 0 & 0 & 0 & 0 & 1 & 1 & 1/4 - P(1, 1, | A_2, B_1)\\
1 & 0 & 1 & 0 & 0 & 0 & 0 & 0 & 1 & 0 & 1 & 0 & 0 & 0 & 0 & 0 & 1/4 - P(0, 0, | A_2, B_2)\\
0 & 1 & 0 & 1 & 0 & 0 & 0 & 0 & 0 & 1 & 0 & 1 & 0 & 0 & 0 & 0 & 1/4 - P(0, 1, | A_2, B_2)\\
0 & 0 & 0 & 0 & 1 & 0 & 1 & 0 & 0 & 0 & 0 & 0 & 1 & 0 & 1 & 0 & 1/4 - P(1, 0, | A_2, B_2)\\
0 & 0 & 0 & 0 & 0 & 1 & 0 & 1 & 0 & 0 & 0 & 0 & 0 & 1 & 0 & 1 & 1/4 - P(1, 1, | A_2, B_2)\\
1 & 1 & 1 & 1 & 1 & 1 & 1 & 1 & 1 & 1 & 1 & 1 & 1 & 1 & 1 & 1 & 0
\end{array}\right], \nonumber \\
c = \left[\begin{array}{ccccccccccccccccc}
0 & 0 & 0 & 0 & 0 & 0 & 0 & 0 & 0 & 0 & 0 & 0 & 0 & 0 & 0 & 0 & 1
\end{array}\right], \nonumber \\
b = \left[\begin{array}{ccccccccccccccccc}
\frac{1}{4} & \frac{1}{4} & \frac{1}{4} & \frac{1}{4} & \frac{1}{4} & \frac{1}{4} & \frac{1}{4} & \frac{1}{4} & \frac{1}{4} & \frac{1}{4} & \frac{1}{4} & \frac{1}{4} & \frac{1}{4} & \frac{1}{4} & \frac{1}{4} & \frac{1}{4} & 1
\end{array}\right], \nonumber \\
x = \left[\begin{array}{ccccccccccccccccc}
p_{lr}(0000) & p_{lr}(0001) & p_{lr}(0010) & \ldots & p_{lr}(1111) & v
\end{array}\right]. \nonumber \\
\end{align}

Let us observe that the above LP depends solely on the set of observables $A_1, A_2, B_1, B_2$ that the observers choose from during the measurements. In our experimental situation any observable can be (and in practice is) effectively parametrised by a pair of angles $\theta, \phi$ so our rightmost probabilities become
\begin{align}
P(r_a, r_b | A_i, B_k) \equiv P(r_a, r_b, \theta^A_i, \phi^A_i, \theta^B_k, \phi^B_k).
\end{align}
The details of why such a parametrization takes place can be found in \cite{mybib:NielsenChuang} so are omitted here. However, what is important is that the critical visibility can be defined as the following function of the angles parameterising the observables.
\begin{align}
v_c(\theta^A_1, \phi^A_1, \theta^A_2, \phi^A_2, \theta^B_1, \phi^B_1, \theta^B_2, \phi^B_2) \in [0, 1]
\label{eq:function}
\end{align}

\subsection{Minimal critical visibility}

A single evaluation of this function is the lower-level task, requiring the solution of the LP problem described above with the maximum of the cost function $z$ as the critical visibility function's value. Our ultimate goal is to minimize~(\ref{eq:function}), thus arriving at the \textbf{minimal critical visibility}, the non-classicality measure that we seek. The minimization of the critical visibility function is the higher-level non-convex non-linear optimization task mentioned in Section~\ref{intro}.

The following sections contain much discussion, both about individual LPs and the higher-level optimization. Hence, we would like to clarify certain vocabulary here. We will from now on use the expression ``solve an LP'' interchangeably with ``calculate the value of the critical visibility function'', and the expression ``minimize the critical visibility function'' interchangeably with ``execute the higher-level optimization procedure''. We will also say ``solve a sequence of LPs leading to the minimum of the critical visibility function'', as this is exactly what the higher-level optimization procedure does. It requests a number of critical visibility function values (each of which requires the solution of a separate LP) in order to minimize it.

This work is mainly concerned with the method of obtaining numerical results rather than the results themselves. Therefore, for simplicity we will only analyse a single quantum entangled state called the Greenberger-Horne-Zeilinger (GHZ) state described in \cite{mybib:GHZ, mybib:GHZ2}.

\subsection{Redundancy in the problem formulation}

One of our first findings was that the LPs constructed as described 
above contain a significant proportion of redundant constraints which increases 
with the problem size. 
For the 65536x65536 problem, where the first number is the number of rows excluding the summation of probabilities and the second is the number of variables excluding the visibility, only 6561 rows are needed for full formulation of the LP problem. We found that, when constructing the matrix $A$, it is possible to skip the rows according to the following rule. If $d$ is the number of possible measurement outcomes (in our case $d=2$), then:

\begin{lemma}
A row is redundant if its last column contains probabilities of measurements for which any of the observers obtained a result equal to $d-1$ while measuring an observable with index higher than $1$. 
\end{lemma}

In our example this means that one can skip rows 6, 8, 11, 12, 14, 15, 16. It is also possible to skip the last row representing the summation of probabilities to one since it is never needed. So, since 8 of 17 rows of matrix $A$ can be removed, the rank of the problem matrix ($A$) is 9. The percentage of redundant rows in matrix $A$ is, therefore, slightly less than 50\%. This percentage, however, increases rapidly with the problem size and allows us to get rid of more than 90\% of the rows for sufficiently large problems. This finding significantly sped up calculations both with the matrix-free interior point method and with the simplex method (GLPK). \\

Let us observe that, typically, both the number of rows and the number
of variables in our LPs are multiples of 2 so, from now on, we will
use `k' to mean 1024, so the 65536x65536 problem will become problem \Problem{64kx64k}. Thus the problem name refers to its origin (since it is
informative for researchers in the field of quantum information science) but the size quoted for the problem is that of the LP after
elimination of redundancy. Furthermore, since all the problems analysed in this paper are square, we will skip one of the dimensions and problem \Problem{64kx64k} will also be referred to as problem \Problem{64k}.

%%%%%%%%%%%%%%%%%%%%%%%%%%%%%%%%%%%%%%%%%%%%%%%%%%%%%%%%%%%%%%%%%%
\section{Original approach}
\label{origapproach}

The original approach adopted by the authors of \cite{mybib:GrucaEtAl} 
used the implementation of the simplex method \cite{mybib:Dantzig} from the GLPK library \cite{mybib:GLPK} for solving the lower-level LP task, and the downhill simplex method 
(DSM) \cite{mybib:NM-DSM} from the SciPy package \cite{mybib:SciPy} as the higher-level 
non-linear optimizer. As shown in \cite{mybib:GrucaB, mybib:GrucaMSc} the critical visibility function is continuous but not differentiable so a robust higher-level optimizer was needed. The DSM is a good fit since, in the vast majority of cases, it can be applied to any continuous function. It is worth mentioning that these two optimization techniques (the simplex method for LP and the DSM)
are extremely powerful tools that have been used successfully in a plethora of applications. 
However, both of them have occasional drawbacks. 
The simplex method may struggle when applied 
to degenerate LP problems \cite{mybib:JH-KM-cycling} and the DSM is not guaranteed to converge to even a local 
minimizer \cite{mybib:Ken}. 
The implementation has since been revised and now uses a newer version
of the GLPK library (4.47 instead of 4.31) and a different
implementation of DSM \cite{mybib:NumRecipes}. This new code is called
\textbf{steam-roller2}. It exploits the property that the LP problems being solved in sequence are related to one another and differ only in the last column of $A$, allowing the simplex solver to hot-start its solution
procedure.

A typical execution of the method chooses random angles as a starting
point for the higher-level DSM optimization procedure. These angles
are used to generate the first LP and its solution yields the value of
the critical visibility function for the starting point. The DSM then
modifies the angles as it sees fit and another LP is then generated
and solved to calculate the critical visibility function value for the
new angles. The procedure thus generates a sequence of LPs, each of which 
is solved by the simplex method.  Finally, after a couple of hundred 
of such steps, the minimum of the continuous non-differentiable critical 
visibility function is found.

Such an approach had a number of advantages: the two algorithms were a good 
fit since the DSM always converged and often returned the global minimum outright,
and the simplex method could take advantage of the similarities between 
the subsequent LPs. They were also very precise: the results could be matched 
with known theoretical values (up to 15 significant digits). This is very much 
the limit of the machine's precision.

It was found, however, that the calculations slowed down when the
problem size increased and the GLPK simplex implementation struggled
to solve the LP problems larger than \Problem{16k}. The simplex
method for linear programming is a non-polynomial algorithm. Indeed,
there does not exist a polynomial bound on the number of iterations
that it may need to solve a given LP and occasionally it may need to
perform a huge number of iterations. Although in practice such
situations are rare \cite{mybib:JH-KM-cycling} the LPs which model the
critical visibility seem to challenge the simplex method. The
solution of a single LP may require hours (or days) of computations
so, since a sequence of hundreds of such LPs may have to be solved,
the whole approach is questionable.

We have made a radical change in the solution approach and decided 
to use an interior point method (IPM) rather then the simplex method 
to solve the LPs. Such is the improvement of efficiency delivered 
by the use of the matrix-free interior point method on these LPs  
that we can afford to sacrifice the hot-start facility of the simplex 
method that would normally dictate its superiority. 
IPMs enjoy a polynomial worst-case iteration complexity. 
Indeed, an LP with $n$ variables is solved to $\varepsilon$-optimality 
in ${\cal O}(\sqrt{n} \log(\frac{1}{\varepsilon}))$ iterations \cite{mybib:Wright}. 
These methods are generally believed to be well-suited to the solution 
of very large scale optimization problems \cite{mybib:G-ipmXXV}. IPMs converge 
to the optimal solution in very few iterations: they usually need only 
 ${\cal O}(\log{n})$ iterations to reach a solution of the problem with $n$ 
variables. However, a single iteration might be costly. A new variant,
called the matrix-free IPM \cite{mybib:G-MtxFree}, removes this 
potential drawback. It replaces the {\it exact} Newton method with 
the {\it inexact} one and employs an iterative method based on conjugate gradients 
to solve the underlying systems of linear equations. In the next section 
we describe briefly the matrix-free interior point method and discuss 
in detail how it is applied to LPs which model the critical visibility. 

\section{New approach}
\label{newapproach}

We start this section with a discussion of a general purpose interior point 
method for linear programming and its special variant called the matrix-free 
interior point method. The reader familiar with IPMs may skip this part. 
The reader interested in more detail on the theory and implementation 
of IPMs should consult the excellent book of Wright \cite{mybib:Wright} and 
the recent survey by Gondzio~\cite{mybib:G-ipmXXV}, respectively. The matrix-free 
IPM itself is introduced by Gondzio~\cite{mybib:G-MtxFree}. 

\subsection{Matrix-free interior point method}
\label{hopdm}

Consider the following linear optimization problem in general form
\begin{equation}
\text{max}\quad c^Tx\quad \text{subject to}\quad Ax=b,\quad x \geqslant 0,
\label{myLP}
\end{equation}
where $A \in R^{m \times n}, x,c \in R^{n}$ and $b \in R^{m}$. An IPM
approaches its solution by moving through the interior of the orthant
defined by the non-negativity constraints. This is achieved by
removing the non-negativity constraints and adding logarithmic barrier
terms $-\mu \ln x_j$ to the objective function to give the problem
\begin{equation}
\text{max}\quad c^Tx -\mu \sum\limits_{j=1}^{n} \ln x_j\quad \text{subject to}\quad Ax=b.
\label{myBarrierLP}
\end{equation}
The parameter $\mu$ controls the force of the logarithmic barrier so
influences the distance of iterates from the boundary of the positive orthant. 
This parameter is gradually reduced as the algorithm progresses, in order 
to allow it to get arbitrarily close to the optimum which, for LP problems,
is always found on the boundary of the orthant \cite{mybib:Wright}. 
The first order optimality conditions for (\ref{myBarrierLP}) are nonlinear 
equations of the form
\begin{eqnarray}
\label{optimalityconditions}
Ax & = & b,\nonumber\\
A^T y + s & = & c,\\
XSe & = & \mu e,\nonumber\\
(x,s) & \geqslant & 0.\nonumber
\end{eqnarray}
Here $y \in R^{m}$ and $s \in R^{n}, s \geqslant 0$ are the dual
variables (Lagrange multipliers) associated with the linear equality
constraint $Ax=b$ and the linear inequalities $x \geq 0$,
respectively, $e \in R^n$ is the vector of ones and $X$ ($S$) is a
diagonal matrix in $R^{n \times n}$ with elements of the vector $x$
($s$) along the diagonal. In interior point methods the system of
equations (\ref{optimalityconditions}) is solved using the Newton
method \cite{mybib:G-MtxFree}. One IPM iteration calculates the Newton
direction, making one step in this direction and reduces the
parameter $\mu$. The Newton direction is obtained by solving
\begin{align}
\label{NewtonEqs}
\left[\begin{array}{cccc}
A & 0 & 0 \\
0 & A^T & I_n\\
S & 0 & X \\
\end{array}\right]
%\cdot
\left[\begin{array}{cccc}
\Delta x \\
\Delta y \\
\Delta s \\
\end{array}\right]
=
\left[\begin{array}{cccc}
 \xi_p \\
 \xi_d \\
 \xi_\mu \\
\end{array}\right]
=
\left[\begin{array}{cccc}
b - Ax \\
c - A^Ty -s \\
\sigma \mu e - XSe \\
\end{array}\right],
\end{align}
where $I_n$ is the identity matrix of size $n$ and $\sigma \in (0,1)$ 
represents the reduction of the parameter $\mu$ in every iteration 
of the algorithm. We can transform (\ref{NewtonEqs}) by eliminating 
$\Delta s = X^{-1}(\xi_{\mu} - S \Delta x)$ to get 
the following symmetric but indefinite augmented system
\begin{align}
\label{AugSys}
\left[\begin{array}{cccc}
-\Theta^{-1} & A^T \\
A & 0 \\
\end{array}\right]
\left[\begin{array}{cccc}
\Delta x \\
\Delta y \\
\end{array}\right]
=
\left[\begin{array}{cccc}
f \\
d \\
\end{array}\right]
=
\left[\begin{array}{cccc}
\xi_d - X^{-1} \xi_\mu \\
\xi_p \\
\end{array}\right],
\end{align}
where $\Theta = X S^{-1}$. 
Further elimination of $\Delta x = \Theta(A^T \Delta y -f)$ gives the
following system of normal equations whose matrix of coefficients is
symmetric and positive definite:
\begin{align}
\label{NEqs}
(A\Theta A^T)\Delta y = g = A \Theta f + d.	
\end{align}
The component-wise products $x_j s_j = \mu$ are driven to zero (recall 
the 3rd equation in (\ref{optimalityconditions})) and ultimately 
define an optimal splitting of indices $j \in \{1, 2, \dots, n \}$ into two 
subsets: $j \in {\cal B}$ for which $x_j = {\cal O}(1)$ and $s_j = {\cal O}(\mu)$ 
and $j \in {\cal N} = \{1, 2, \dots, n \}\setminus{\cal B}$ for which 
$x_j = {\cal O}(\mu)$ and $s_j = {\cal O}(1)$. Consequently, the diagonal 
scaling matrix $\Theta$ is very ill-conditioned: when $\mu$ approaches 
zero, the elements $\Theta_j$ go to infinity or zero for indices 
$j \in {\cal B}$ or $j \in {\cal N}$, respectively. The matrices 
of linear systems (\ref{AugSys}) and (\ref{NEqs}) are therefore both very 
ill-conditioned. Different regularization techniques have been proposed 
to cure this ill-conditioning. Following Saunders \cite{mybib:S-pdReg}, 
Altman and Gondzio \cite{mybib:AG-QP} propose regularizing both the primal 
and the dual formulation of the problem, replacing (\ref{AugSys}) by
\begin{align}
\label{regAugSys}
\left[\begin{array}{cccc}
-(\Theta^{-1} + R_p) & A^T \\
A & R_d \\
\end{array}\right]
\left[\begin{array}{cccc}
\Delta x \\
\Delta y \\
\end{array}\right]
=
\left[\begin{array}{cccc}
f^\prime \\
d^\prime \\
\end{array}\right].
\end{align}
Here $R_p \in R^{n \times n}$ and $R_d \in R^{m \times m}$ are the dynamically 
chosen primal and dual positive definite diagonal regularization matrices and 
$f^\prime \in R^n$ and $d^\prime \in R^m$ are appropriately computed right 
hand side vectors. 
The regularized version of the normal equations is obtained by pivoting 
on the $(1,1)$ block of (\ref{regAugSys}) to give
\begin{align}
\label{regNEqs}
(A(\Theta^{-1} + R_p)^{-1}A^{T} + R_d)\Delta y = g = A(\Theta^{-1}+R_p)^{-1}f' + d^\prime,
\end{align}
Thus the matrix $G = A \Theta A^{T}$ in (\ref{NEqs}) is replaced by
$G_R = A(\Theta^{-1} + R_p)^{-1}A^{T} + R_d$ in (\ref{regNEqs}). 

The matrix-free interior point method \cite{mybib:G-MtxFree} applies 
the {\it inexact} Newton method to (\ref{NewtonEqs}). It replaces 
(\ref{NEqs}) by the regularized system (\ref{regNEqs}) and applies the preconditioned 
conjugate gradients algorithm to solve (\ref{regNEqs}) only approximately, and
to rather loose accuracy. The resulting solution $\Delta y$ is used 
to compute the corresponding $\Delta x$ and $\Delta s$, but clearly 
the complete Newton direction $(\Delta x, \Delta y, \Delta s)$ is only 
an approximate solution to (\ref{NewtonEqs}). It is an {\it inexact Newton direction}
for the first order optimality conditions (\ref{optimalityconditions}). 

To accelerate the convergence of the conjugate gradients method the system is preconditioned by a matrix $P$ to reduce its condition number $\kappa$ so that
\begin{align}
\kappa(P^{-1}G_R) \ll \kappa(G_R).
\end{align}
For matrix-free IPM, a specially designed preconditioner
\cite{mybib:G-MtxFree} is identified as follows. Consider the rank $k$
partial Cholesky decomposition% of $G_R$
\begin{align}
\label{PartialCol}
G_R = 
\left[\begin{array}{cccc}
L_{11} & \\
L_{21} & I \\
\end{array}\right]
\left[\begin{array}{cccc}
D_{L} & \\
 & S \\
\end{array}\right]
\left[\begin{array}{cccc}
L_{11}^T & L_{21}^T \\
& I \\
\end{array}\right],
\end{align}
where $S \in R^{(m-k)\times (m-k)}$ is the Schur complement obtained by eliminating 
$k$ pivots, $L = [L_{11}^T ~ L_{21}^T]^T$ is a trapezoidal matrix containing the first 
$k$ columns of the Cholesky factor of $G_R$ and $D_L \in R^{k \times k}$ is a diagonal 
matrix containing the $k$ largest pivots of $G_R$. The preconditioner
\begin{align}
\label{Prec}
P = 
\left[\begin{array}{cccc}
L_{11} & \\
L_{21} & I \\
\end{array}\right]
\left[\begin{array}{cccc}
D_{L} & \\
 & D_{S} \\
\end{array}\right]
\left[\begin{array}{cccc}
L_{11}^T & L_{21}^T \\
& I \\
\end{array}\right]
\end{align}
is~(\ref{PartialCol}) with the Schur complement matrix $S$ replaced by its diagonal $D_S$. 

The rank $k$ of the partial Cholesky decomposition is carefully chosen
to obtain an optimal trade-off between precision and execution
time. In general the approximation becomes more precise as the rank
increases, but greater effort has to go into its computation and use. When
$k=m$ the Cholesky decomposition is complete: there is no Schur
complement matrix and $P^{-1}G_R=I$ so the conjugate gradients method
terminates in one iteration. However, for reasons of efficiency, we
always set $k \ll m$. The choice of rank of the partial Cholesky used
in the LP problems considered in this paper will be discussed in the
following sections.

Another important feature of the matrix-free interior point method is its 
ability to allow for an implicit treatment of matrix $A$ 
\cite{mybib:G-MtxFree}. This is relevant when $A$ is very large 
and therefore requires a lot of memory to store. This is clearly the case 
for LPs arising in quantum information problems. The matrix-free IPM needs 
only the following {\em results\/} of operations performed with matrix $A$.
\begin{enumerate}
\item[(i)] 
Multiply the matrix $A$ and vector $x$.
\item[(ii)] 
Multiply the matrix $A^T$ and vector $y$.
\item[(iii)] 
Retrieve an arbitrary column of the matrix $A\Theta A^T$.
\item[(iv)] 
Retrieve the diagonal of the matrix $A\Theta A^T$.
\end{enumerate}

Therefore, neither matrix $A$ nor matrix $G_R$ needs 
to be fully stored in order to solve (\ref{regNEqs}). All that is needed 
is the implementation of the above operations. In the following sections 
we will discuss the implications of this feature and we will illustrate 
the practical behaviour of the method both in terms of the memory and 
CPU time requirements.

\subsection{Implementation}

As described above the matrix-free interior point method requires 
implementations of a number of operations which define the LP. 
Here is a full summary:
\begin{enumerate}
\item[(i)]   \FSAX:  the product of the matrix $A$ and vector $x$, which is a parameter,
\item[(ii)]  \FSATY: the product of the matrix $A^T$ and vector $y$, which is a parameter,
\item[(iii)] \FCAAT: the $i$-th column of the matrix $A\Theta A^T$, where $i$ is a parameter,
\item[(iv)]  \FDAAT: the diagonal of the matrix $A\Theta A^T$.
\end{enumerate}

We implemented them all in a program integrating \textbf{steam-roller2} 
and the matrix-free interior point method of \textbf{HOPDM} software 
\cite{mybib:G-MtxFree}. 

\subsubsection{Performance optimization}

The above functions are called very frequently during program execution 
and are responsible for a significant percentage of the CPU time used. 
Hence, it is essential that they are implemented efficiently. 
Our implementations are based on a row-wise storage of the matrix $A$. 
For each row only the column indices of non-zeros are stored 
to save space (with the exception of the last column of the matrix, 
for which real values in the interval $[0,1]$ also need to be stored).
This turned out to be a very efficient method of storage, 
with a \Problem{64k} problem consuming less than 10 MB of memory. 
Indeed, after the redundancy of the problem has been removed, 
the memory required to store the problem grows sub-linearly with 
its dimension. 
There are two reasons for this:
\begin{enumerate}
\item The vast majority of nonzero entries are equal to 1.
\item The percentage of redundant constraints increases with the problem size.
\end{enumerate}
Our first implementations contained solely the definition of how each
of the operations should be performed in the C programming language
(without any optimization) and showed how much room for optimization
there was. All results in this section are obtained from a development
workstation and are, therefore, non-indicative in terms of absolute
times and a production code would be much faster (see next section for
production code results).  Table~\ref{tab:single1} contains output
from the UNIX operating system's profiler \gprof when solving a single
\Problem{16k} problem.  After reduction of redundancy we have $m=2187$
and $n=16385$, the latter being $16 \times 1024$ plus one variable
representing the visibility. We report the number of calls of each
routine, the CPU time (in seconds) per 1000 calls, and the percentage
of overall solution time spent while executing a given routine.

\begin{table}[!htc]
\centering
\begin{tabular}{lrcc}
\toprule
operation & calls & time / 1000 calls \ & \ \% of total time \\
\midrule
\FSAX  & 12306 & 1.255 & 13.22 \\
\FSATY & 11307 & 3.766 & 36.46 \\
\FCAAT &  1000 & $\approx$ 0 & $\approx$ 0 \\
\FDAAT &    10 & $\approx$ 0 & $\approx$ 0 \\
\bottomrule
\end{tabular}
\caption{Execution time profile for a single \Problem{16k} LP with unoptimized functions.}
\label{tab:single1}
\end{table}

Only the costs of \FSAX and \FSATY are significant, accounting for about
half of the execution time, so the following optimizations were
performed to improve their efficiency.

{\em Optimization of \FSAX.} Since all columns of $A$ apart from the
last contain either ones or zeros, the contributions of these entries
in a summation $Ax=u$ are either skipped (if zero) or added (if one)
when the intermediary sum builds a particular component of the
resulting vector. Thus the multiplication by one that would occur in
general is skipped. For sufficiently large problems (larger than
\Problem{256}) loop unrolling is performed (following an established
practice described, for example, by Sarkar and
Vivek~\cite{mybib:Sarkar} and commonly used in, for example,
LAPACK~\cite{mybib:LAPACK}). Each row contains the same number of
non-zero elements. These are divided into groups of 16 and added to
the intermediate sum all at once. This reduces the loop control
overhead and yields an almost two-fold speed-up of the \FSAX routine
(see Table~\ref{tab:single2}).\\

\begin{table}[!htc]
\centering
\begin{tabular}{lrrr}
\toprule
operation           & calls & time / 1000 calls \ & \ \% of total time \\
\midrule
\FSAX    & 11943 & 0.677 &  9.07 \\
\FSATY   & 10944 & 2.404 & 29.50 \\
\bottomrule
\end{tabular}
\caption{Execution time profile for a single \Problem{16k} LP with loop unrolling.}
\label{tab:single2}
\end{table}

{\em Optimization of \FSATY.} As for \FSAX, 
multiplication by one is skipped and loop unrolling is performed. 
Here it is somewhat less efficient due to the row-wise storage 
of matrix $A$ (see Table \ref{tab:single2}). The intermediate sums 
have to be stored in the resulting $n$-dimensional vector itself and, despite 
loop unrolling, they are frequently accessed at each step of the algorithm. 
Further speed-up was achieved when column-wise storage of matrix $A$ 
was used. This enabled us to use intermediary sums and store them only once 
in the resulting vector, just as in the case of \FSAX. 
As mentioned at the beginning of this section, storage is very efficient 
so we can safely afford to store a duplicate copy of the problem matrix. 
The implementation of loop unrolling for the column-wise 
storage was more difficult due to the different number of nonzero 
elements in columns, with even very large problems containing multiple 
columns with one, three or five elements. After a considerable effort, 
resulting in more than a hundred lines of loop unrolling code, we achieved 
the significant speed-up seen in Table \ref{tab:single3}. 
The total resulting speed-up of \FSATY after optimizations 
is more than three-fold compared to unoptimized version of the function.\\

\begin{table}[!htc]
\centering
\begin{tabular}{lrcc}
\toprule
operation           & calls & time / 1000 calls \ & \ \% of total time \\
\midrule
\FSAX    & 11697 & 0.694 & 11.19 \\
\FSATY   & 10698 & 0.955 & 14.08 \\
\bottomrule
\end{tabular}
\caption{Execution time profile for a single \Problem{16k} LP (\FSATY uses 
         the column-wise storage of $A$ and an optimized loop unrolling).}
\label{tab:single3}
\end{table}

We do not provide total execution times as they are not informative. 
This is due to the order of additions and the associated numerical errors, 
causing the optimized and unoptimized implementations to differ 
in the last significant digits and, hence, gradually alter the path to optimality 
taken by the matrix-free interior point method. This is also the reason 
why the optimized and the unoptimized implementations have different 
numbers of function calls. Having run the program many times for different 
input we noticed a significant variation in the number of calls between  
the optimized and unoptimized functions and it is safe to assume that 
the difference does not favour either implementation. Neither of the two versions is numerically superior to the other in terms 
of precision and they would behave identically if no numerical errors were present. 

Significant overall speed-up of execution in the range 25--75\% was observed 
after implementing the above improvements.

\subsubsection{Precision}
\label{precision}

Obtaining sufficient precision was the major challenge of our
investigations. It depends on a number of parameter settings and we
have made a considerable effort to choose these parameters
optimally. In the results reported below we analyse problems with
$d=2$ so there are two possible outcomes for any measurement ($0$ or
$1$), 2 observables per each observer and increasing numbers of
observers. In this setting the problem size grows four-fold each time
an observer is added to the problem. So the minimal problem with 2
observers is \Problem{16}, the problem with 3 observers is
\Problem{64}. However, after eliminating redundancy, the size of LP
increases three-fold with each additional observer.

The rank of partial Cholesky determines precision since, when it is
large, it allows for more precise direction calculations. It cannot,
however, be increased indefinitely lest the execution time exceed that
of the simplex-based approach. We found a value of 100 to be optimal
for our computations and we used it for all problem sizes. As a
result, for LP problems of size 100 or less the preconditioner is
exact so conjugate gradients is exact in one iteration and the
interior point method uses the exact Newton direction. Since problems
\Problem{256} and \Problem{1k} (with 4 and 5 observers respectively)
yield LPs of size 81 and 243 respectively, we only analyse problems
with at least 5 observers since only for these problems is conjugate
gradients non-trivial.

The matrix-free interior point method applies the preconditioned conjugate 
gradients algorithm to the system (\ref{regNEqs}) and uses the partial 
Cholesky preconditioner (\ref{Prec}). The solution of (\ref{regNEqs}) 
is controlled by two parameters: the maximum number of conjugate gradient 
iterations allowed and the desired precision achieved by the algorithm.
The use of high accuracy and/or setting a large conjugate gradients 
iterations limit might increase the execution time but these settings 
are needed to provide the accuracy required by (inexact) Newton directions. 
We vary these settings and start from a lower iteration limit at the first 
steps of IPM, gradually increasing it to 1000 towards 
the end of optimization. We also set the relative error tolerance 
in the CG algorithm to $\varepsilon_{CG} = 10^{-6}$. 

Another parameter which influences the precision of the method is the 
optimality tolerance $\epsilon$ used to terminate the IPM when
\begin{align}
   \frac{|c^Tx - b^Ty|}{|c^Tx| + 1} \leq \epsilon, 
\end{align}
where $c, b, x, y$ are defined in Section~\ref{hopdm}. 
We set $\epsilon = 10^{-5}$ for all problems. \\

With the above settings we managed to obtain two-digit exact optimal 
solutions for all problems up to \Problem{16k} that were attempted. 
We also achieved a precision of $\pm 25\%$ in the solutions for larger 
problems. Although these solutions are less accurate than those published 
in \cite{mybib:GrucaEtAl} they are sufficiently accurate to elicit 
the qualitatively important properties of most states. In practice 
the results are typically much more precise than this worst-case value, 
as may be seen in the tables in Section~\ref{perfcomparison}.

%%%%%%%%%%%%%%%%%%%%%%%%%%%%%%%%%%%%%%%%%%%%%%%%%%%%%%%%%%%%%%%%%
\section{Performance comparison}
\label{perfcomparison}

We begin our comparison with individual LPs solved by the two
approaches: the simplex method (GLPK) and the matrix-free interior
point method (HOPDM). Results in this section are based on runs on a
computer with an {\tt Intel\textsuperscript{\textregistered}
Core\texttrademark\ i7 3.07GHz} processor and 24 GB of \RAMm. 
Computational tests are performed using a Bell experiment with 2
observables per observer since the minimal critical visibility 
of the GHZ state is known and given in~\cite{mybib:obydwa} as
\begin{align}
\label{eq:MinCritVis}
\min v_{c} = 2^{\frac{1 - n}{2}},
\end{align}
where $n$ is the number of observers. 

Table~\ref{tab:statistics} reports the worst-case numerical error for
results obtained
with the matrix-free interior point method during an execution of the
program which stopped after one hundred critical visibility function
evaluations. This means that one hundred LPs were generated and solved
successively. The angles used to generate LPs were provided by the
higher-level optimization method (DSM), as in a typical execution of
the program. For a given number of observers, each line in
Table~\ref{tab:statistics} reports different execution results, one
for each problem size. The reduced problem size corresponds to
the number of rows in the constraint matrix after the redundancy is
removed. For all problems solved with both solvers (GLPK and HOPDM)
the solution results have been compared. The last column in the table
contains the largest difference between the objective returned by GLPK
(which have been confirmed repeatedly to be correct and precise) and
by matrix-free HOPDM.

\begin{table}[!htc]
\centering
\begin{tabular}{lrrr}
\toprule
observers & problem name & reduced problem size & the largest difference \\
\midrule
5 & \Problem{1k}  &  243 & 0.006535 \\
6 & \Problem{4k}  &  729 & 0.006738 \\
7 & \Problem{16k} & 2187 & 0.003662 \\
\bottomrule
\end{tabular}
\caption{Worst-case numerical error for 100 consecutive LPs of each problem size.}
\label{tab:statistics}
\end{table}

In Table~\ref{tab:performance1} we report the time required by the two
solvers, the expected solution and the optimal objective function
value (critical visibility) returned by HOPDM. The value returned by
GLPK always matched the expected solution obtained analytically so it
is skipped from the tables. Values calculated with~(\ref{eq:MinCritVis})
are given in the solution column of the table. The angles used to
generate each LP were the same for both GLPK and HOPDM. They were also
minimal for each problem for the GHZ state. Each line in
Table~\ref{tab:performance1} corresponds to one LP. Problems
\Problem{64k} and larger were scaled so that the returned value was in
fact 8 times greater. This helped preserve the precision of the
results, where the value of the solution is close to zero. As one can
see from studying the table some of them are still off by a small
fraction, which is the expected speed-accuracy trade-off.

\begin{table}[!htc]
\centering
\begin{tabular}{r|rr|c|c|cc}
\toprule
& \multicolumn{2}{c}{problem} & & \specialcell{Simplex\\ Method} & \multicolumn{2}{c}{\specialcell{Interior Point\\ Method}} \\
\midrule
$n$ &name &  size & solution  & time & time & solution \\
\midrule
4 & \Problem{256} &    81  &     0.354  & $\approx$ 0s & $\approx$ 0s  &  0.354 \\
5 & \Problem{1k}  &   243  &     0.250  &        0.02s &  0.08s  &  0.250 \\
6 & \Problem{4k}  &   729  &     0.177  &        0.93s &  0.87s  &  0.177 \\
7 & \Problem{16k} &  2187  &     0.125  &        1m13s &  11.88s &  0.125 \\
8 & \Problem{64k} &  6561  &     0.088  &        6h51m &  3m22s  &  0.088 \\
9 & \Problem{256k}& 19683  &     0.063  &    $\gg$ 24h &  28m38s &  0.068 \\
10 & \Problem{1m} & 59049  &     0.044  &      unknown &  1h34m  &  0.051 \\
11 & \Problem{4m} & 177147 &     0.031  &      unknown &  9h15m  &  0.029 \\
\bottomrule
\end{tabular}
\caption{Comparison of the performance of the simplex method (GLPK)
and interior point method (HOPDM) applied to solve a single LP. The
$n$ column represents the number of observers in a Bell experiment
with 2 observables per observer. Minimal angles for the GHZ state were
used in each case.  The solution column contains results obtained
analytically. The results returned by the simplex method always
matched the analytical values so they are omitted.}
\label{tab:performance1}
\end{table}

The analysis of results collected in Table~\ref{tab:performance1}
reveals that the simplex method is faster for smaller problems but its
solution time grows very quickly with problem size. For problem
\Problem{16k} (of size 2187) GLPK is already an order of magnitude
slower than HOPDM.

Computations using GLPK for problem \Problem{256k} were terminated after about 
24 hours and computations for problem \Problem{1m} were not even 
attempted as they cannot be expected to complete within any reasonable time. 
HOPDM times, on the other hand, increase roughly linearly with problem size, 
with successive factors~\label{sequence} being 10.9, 13.66, 17.00, 8.50, 3.29, 5.88. 
This confirms that there is no explosion of algorithm steps to take when 
solving the LP as the problem size increases. One can expect that 
the computation time will increase linearly with reduced problem size
(and depend very little on the number of variables in the problem).

It should be noted, however, that the simplex method has a significant
advantage over the interior point method when a sequence of similar
LPs are solved. Successive LPs differ merely by the last column in the
matrix $A$ which contains the angles and therefore one may expect that
the solutions of two successive problems are similar. The simplex
method in GLPK has an option which allows the optimal solution of one
problem to be used to warm-start another and this technique proved
very useful. Although, in general, it is possible to use the
warm-starting facility of IPMs, this option is not available for the
matrix-free variant of IPM. Therefore, another comparison which
reflects this relative advantage of the simplex method is needed. The
results collected in Table~\ref{tab:performance2} demonstrate the
performance of the complete approach which uses DSM to solve the upper
level optimization problem and employs either the simplex method
(GLPK) or the matrix-free interior point method (HOPDM) to solve the
LPs at the lower level. For both variants of the program we report the
number of LPs solved, the overall time of running the program and the
optimal function value. We report only two significant digits of the
result which corresponds to the precision requested from the DSM
procedure. 
\begin{table}[!htc]
\centering
\begin{tabular}{r|rr|c|rr|rrc}
\toprule
&  \multicolumn{2}{c}{problem} & & \multicolumn{2}{c}{Simplex Method} & \multicolumn{3}{|c}{Interior Point Method} \\
\midrule
$n$ & name &  size & solution & \# LPs &  time & \# LPs &  time & solution \\
\midrule
2 & \Problem{16}    &     9 &     0.71 &        44 & $\approx$ 0s  &        44   &   0.05s  & 0.71 \\
3 & \Problem{64}    &    27 &     0.50 &        78 & $\approx$ 0s  &        78   &   0.13s  & 0.50 \\
4 & \Problem{256}   &    81 &     0.35 &       156 &        0.07s  &       156   &   0.90s  & 0.35 \\
5 & \Problem{1k}    &   243 &     0.25 &       231 &        1.67s  &       161   &     21s  & 0.25 \\
6 & \Problem{4k}    &   729 &     0.18 &       251 &       	  56s  &       265   &   5m28s  & 0.18 \\
7 & \Problem{16k}   &  2187 &     0.13 &       346 &        2h00m  &       330   &  56m52s  & 0.13 \\
8 & \Problem{64k}   &  6561 &     0.088 &  unknown &    $\gg$ 24h  &       331   &  16h47m  & 0.090 \\
\bottomrule
\end{tabular}
\caption{Efficiency comparison of two approaches for computing the minimal critical visibility of the GHZ state. Subsequent LPs are solved with the simplex method (GLPK) and the matrix-free IPM (HOPDM). The $n$ column represents the number of observers in a Bell experiment with 2 observables per observer. The solution column contains results obtained analytically. The results returned by the simplex method always matched the analytical ones so they are skipped from the table. The last row of the table uses adjusted settings for both the DSM procedure and HOPDM --- see main text (page \pageref{adjustments:8}).}
\label{tab:performance2}
\end{table}

In both cases the same set of angles (drawn randomly beforehand) were provided 
to DSM as a starting point for the optimization process. Since GLPK and HOPDM 
return slightly different results for a single LP, the execution of DSM differs 
slightly (as one can deduce from the different number of LPs generated). 
Both, however, return the same correct result to within two-digits, except for problem 64k.

For the three smallest problems the number of LPs solved by both
approaches are the same. This is because the number of constraints for
first three problems in our test set (9, 27 and 81) are less than the
rank of the partial Cholesky preconditioner so the latter is the
complete Cholesky decomposition of $G_R$. This perfect preconditioner
allows the matrix-free IPM to produce very accurate optimal
solutions. Analysis of results collected in
Table~\ref{tab:performance2} confirms an advantage of the simplex
method approach over the matrix-free interior point method for smaller
problems. However, it also indicates that for problem \Problem{16k}
the latter approach provides better efficiency.

The largest problem for which it was possible to obtain the solution
with the simplex method was 16k. Larger problems are intractable by GLPK due to
time constraints.
It was possible to obtain a solution to a
problem 4 times larger, that is problem of size 64k, with the IPM method.
In order to obtain a reliable solution we had to make two adjustments \label{adjustments:8} to the computation method. Firstly, we had to relax the sensitivity of the DSM procedure to numerical errors of the optimized critical visibility function. This adjustment was necessary as the cumulative precision error introduced by inaccurate solutions to LPs misguides the DSM procedure and may eventually hamper its convergence. We set the $ftol$ parameter in our DSM implementation to $0.04$ compared to a previous setting of $0.01$. $ftol$ is the fractional convergence tolerance, which we have found to return results correct on the second decimal place for the standard setting of $0.01$ used for all results in this paper so far. After relaxing this parameter by setting it to $0.04$ we can expect a less precise result, that is one potentially above the minimum, but also easier convergence. For details see \cite{mybib:NumRecipes}. Having made this one adjustment we obtained a result $0.094$, compared to the correct result of $0.088$. We obtained a further improvement of the computational result by increasing the size of the partial Cholesky preconditioner rank from $100$ to $150$. The result of $0.090$ obtained this way is reported in Table~\ref{tab:performance2}. This last problem raises a question of whether such a result, which is not precisely equal to the analytical predictions, is of use to the quantum information researcher. There are two reasons why this is so. The first one is that we can have confidence in this result. The inaccuracy comes from the lower-level optimization procedure feeding results to the higher level optimization task. Despite this inaccuracy the higher level optimization procedure did converge, which means that the result fulfilled all necessary (relaxed) convergence conditions. This indicates that the inaccuracy was kept within certain bounds throughout the optimization process. We can also have confidence in the result as due to the nature of the optimization process this final result is vitiated by the error of only the last LP computation. This means that the errors do not propagate in between the subsequent calls of the higher level optimization procedure. It is possible that, due to inaccuracy of the interior point method, the DSM procedure will fail to converge to a global minimum (but then again there is no guarantee it will converge to a global minimum in any case \cite{mybib:Ken}). It is not possible, however, that due to the IPM inaccuracy DSM will converge to a point below the global minimum. In other words, the lowest result we can get will occur in a situation where we got a global minimum (point defined by angles), and for this point we got a critical visibility vitiated by the error of only the final LP calculation. For a researcher in the quantum information domain such a result can be harnessed in the following way. First, one would execute the full DSM optimization with the IPM method to arrive quickly at the global minimum (or somewhere nearby). Second, one would execute a single LP calculation using this point and a more precise LP method like the simplex method to confirm the result. If this is not possible for some reason, then at least the following conclusion can be made. The result obtained might be arbitrarily higher than the desired minimal critical visibility, depending on the convergence of the DSM procedure. But if it is lower, then it will not be by more than the inaccuracy of the single last LP calculation made. That is, the preceding LP calculation errors do not affect the final result in any way other than hampering the convergence of DSM. This allows the quantum information researcher to compare the non-classicality of quantum entangled states to within the error bound introduced solely by the final LP computation. The above example suggests that in the case of problem size 64k the researcher can compare two states to within 0.002, which in many cases allows to elicit qualitative differences between states. It is worth adding that to protect ourselves from the risk of not getting the global minimum as a result of the DSM computation, we repeat it a number of times, be it with the simplex method or the IPM method as the underlying LP solver.

Let us highlight two other important points related to
Table~\ref{tab:performance2}. Firstly, as would be clear to any researcher in the
quantum information domain, there is a world of investigation
opportunities available for problems of sizes between 16k and 64k (and potentially
also slightly larger). It is possible to create problems of sizes in these bounds
by adjusting the number of observers in an experiment (for simplicity this was
the only variable in this section of this paper), the number of outcomes of
any single measurement (which we assumed to be always 2), and the number
of observables available for selection by any of the observers. The last of these
was assumed to be always 2, whereas in practice not only can it vary for
different experiments --- it can also vary between different observers in
the same experiment. With so many variables one can generate problems of almost
any size in the range 16k -- 64k. The majority of these problems were beyond
computational reach before the introduction of the interior point method in the calculations.
All of them can now be solved by HOPDM by making the above adjustments.

Secondly, the interior point method returns reasonably precise results for any
single LP, but the inconsistencies involved are random leading to DSM convergence problems.
Thus, global optimization
does not demonstrate the biggest benefit of using HOPDM. The real power of
the interior point method lies
in single LP calculations, where the slight numerical error doesn't impact the
general contribution of the result. Deducing potential minimal angles is
possible and frequently applied \cite{mybib:obydwa}. Therefore, even when full DSM minimization is not possible as in the case of problems larger than \Problem{64k}, single LP calculations are still of paramount value 
for researchers in the field as they are a way to calculate the critical visibility for the deduced minimal angles. There is a substantial range of problems
made available for numerical investigations thanks to the interior
point method --- see Table~\ref{tab:performance1} and the next section for details.

\begin{center}
\begin{figure}
\begin{center}
\includegraphics{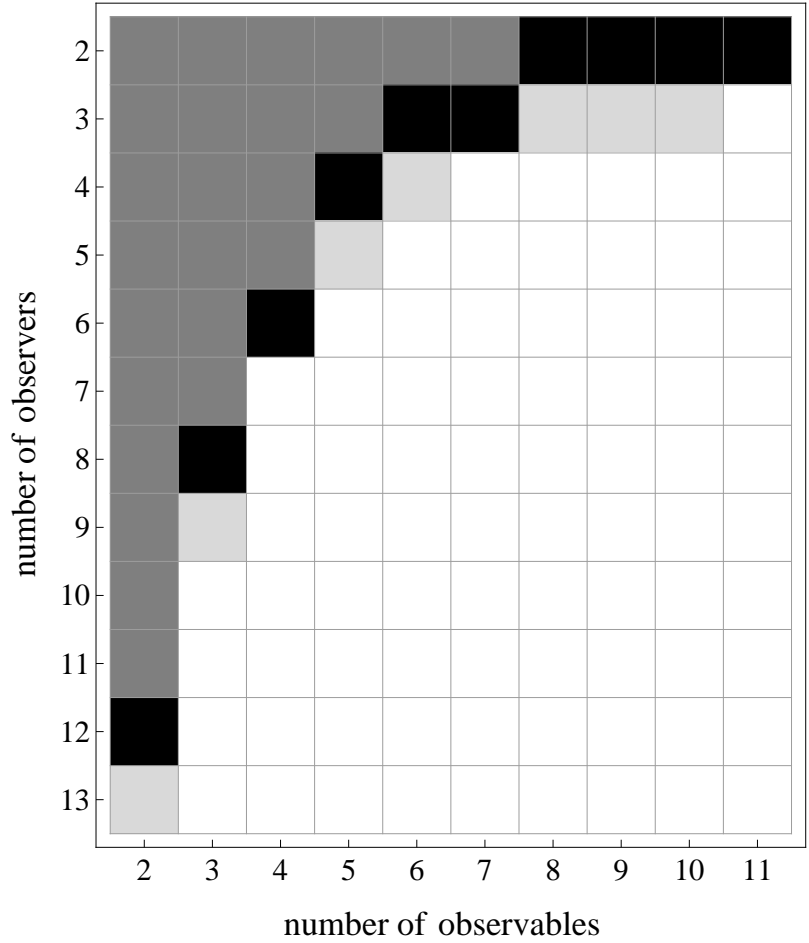}
\caption{Comparison of the range of single-LP problems possible to solve with the simplex method and the interior point method in 24 hours or less. Gray indicates problems possible to solvable with both the interior point and the simplex method, black indicates problems solvable with the interior point method only. Light-grey indicates problems, which could potentially be solved with the interior point method, if sufficient \RAMm was available. For each square the same number of observables was available to all observers in the experiment and is given on the ``number of observables'' axis.}
\label{fig:range}
\end{center}
\end{figure}
\end{center}

\section{Comparison of range of problems within reach}
\label{range}

Finally let us present a broader range of single-LP problems that it
is possible to solve with the simplex method and the interior point
method. Table \ref{tab:range} and Figure \ref{fig:range} summarize
this comparison. For each count of observers we solved the minimal
problem and steadily increased the number of observables available to each of
the observers by one (that is, for 2 observers we began with a 2x2
observable configuration, then tried 3x3, 4x4 and so on). We continued
this process until we encountered a problem which either used too much
\RAMm or exceed a 24 hour computation time threshold. We selected 24
hours as a threshold as we found that this is a maximal period which
allows comfortable trial and error work with different states (used
e.g. in \cite{mybib:GrucaEtAl}).

\begin{table}[!htc]
\centering
\begin{tabular}{r|c|c|r|c}
\toprule
& Simplex Method & \multicolumn{3}{c}{Interior Point Method} \\
\midrule
$n$ & \specialcell{Maximal \#\\of observables} & \specialcell{Maximal \#\\of observables} & \specialcell{Time\\required} & \specialcell{Potentially \\within reach\\with more RAM}\\
\midrule
2 & 11x11 		& 12x12  		&    22m & 13x13 \\
3 & 7x7x7 		& 8x8x8  		&  1h15m & 9x9x9 \\
4 & 5x5x5x5 		& 6x6x6x6 		& 22h11m & - \\
5 & 3x3x3x3x3 		& 4x4x4x4x4 		&  2h00m & 5x5x5x5x5 \\
6 & 2x2x2x2x2x2 	& 3x3x3x3x3x3 		&    41m & 4x4x4x4x4x4 \\
7 & 2x2x2x2x2x2x2 	& 3x3x3x3x3x3x3 	&  9h23m & - \\
8 & 2x2x2x2x2x2x2x2	& 2x2x2x2x2x2x2x2 	&     3m & 3x3x3x3x3x3x3x3 \\
9 & - 			& 2x2x2x2x2x2x2x2x2  	&    28m & 3x3x3x3x3x3x3x3x3 \\
10 & - 			& 2x2x2x2x2x2x2x2x2x2 	&  1h34m & 3x3x3x3x3x3x3x3x3x3  \\
11 & -			& 2x2x2x2x2x2x2x2x2x2x2 &  9h15m & - \\
\bottomrule
\end{tabular}
\caption{Comparison of the range of single-LP problems possible to solve with the simplex method and the interior point method in 24 hours. The ``Potentially within reach with more RAM'' column lists those problems, for which the preceding problem's solution time was 2 hours or less.}
\label{tab:range}
\end{table}

As one can see in both Table \ref{tab:range} and Figure
\ref{fig:range}, for each count of observers experiments with greater
number of observables are within reach of the interior point
method. By ``within reach'' we mean that it is possible to compute
them on our production machine in 24 hours or less. Also, a greater
number of observers is within reach of the interior point method than
of the simplex method in the case of the minimal number 2 of
observables. It should also be noted that, as with all other results
presented in this section, we were limited to the 24 GB of \RAMm
available on the machine we used for the computations. Thus, a number
of problems can be expected to complete in a reasonable time, but it
was not possible to verify it due to limited RAM. We indicated such
problems in both the table and the figure. Our rule of thumb for
distinguishing problems potentially possible to solve from
the ones definitely beyond our reach, even with enough \RAMm, is very
approximate and roughly based on the sequence of time-increase factors
(see Table~\ref{tab:performance1}). We assumed that if the preceding
problem can be solved in about 2 hours or less, then the next problem
could potentially be solved in 24 hours or less. Our estimation is
obviously not precise and the results in the right-most column of the
table should be treated as directions of future investigation, when
machines with more \RAMm are available.

\section{Conclusions}
\label{conclusions}

Minimal critical visibility of a quantum entangled state is a very useful measure in quantum information. However its computation poses a substantial numerical challenge. At the heart of this challenge lie the very large linear programming problems, the solution of which is necessary each time the critical visibility itself is calculated. Previous work always used the simplex method to solve these LPs, an approach providing satisfactory precision in the computed results, but restricting researchers to a limited range of problems which can be solved. We proposed to replace the simplex method with an advanced matrix-free interior point method HOPDM and found that it enabled us to analyse problems much faster and extend our investigations to more complicated ones, which were not previously within our reach. We found that even though the results provided by the interior point method were less precise, they sufficed to analyse, order and compare quantum entangled states, even in complicated cases. 

\section{Acknowledgements}

This work was carried out under the HPC-Europa2 project (project number: \mbox{228398}) with the support of the European Commission Capacities Area-Research Infrastructures Initiative. This work is also available as Technical Report \mbox{ERGO-2012-004} (for other papers in this series see \mbox{\url{http://www.maths.ed.ac.uk/ERGO/}}). Jacek Gondzio was supported by EPSRC grant \mbox{EP/I017127/1}. Wies{\l}aw Laskowski and Marek \.{Z}ukowski are supported by a MNiSW Grant \mbox{IdP2011 00361} (Ideas Plus).

%%%%%%%%%%%%%%%%%%%%%%%%%%%%%%%%%%%%%%%%%%%%%

\bibliographystyle{model1-num-names}
\bibliography{jgjg}

\end{document}